\documentclass[epj-spec]{svjour}
\usepackage{graphics}
\usepackage{dcolumn}
\usepackage{bm}
\newcommand{\bra}[1]{\langle#1|}
\newcommand{\ave}[1]{\langle#1\rangle}

\newcommand{\scal}[2]{\langle#1|#2\rangle}
\newcommand{\ket}[1]{|#1\rangle}
\begin{document}
\title{Correspondence between field generalized binomial states and coherent atomic states}
\subtitle{}

\author{Rosario Lo Franco\thanks{\email{lofranco@fisica.unipa.it}}\and Giuseppe Compagno\and Antonino Messina\and Anna Napoli}
\institute{Dipartimento di Scienze Fisiche ed Astronomiche,
Universit\`{a} di Palermo, via Archirafi 36, 90123 Palermo, Italy}
\abstract{We show that the $N$-photon generalized binomial states of electromagnetic field may be put in a bijective mapping with the coherent atomic states of $N$ two-level atoms. We exploit this correspondence to simply obtain both known and new properties of the $N$-photon generalized binomial states. In particular, an over-complete basis of these binomial states and an orthonormal basis are obtained. Finally, the squeezing properties of generalized binomial state are analyzed.}
\maketitle
\section{\label{intro}Introduction}
Binomial states (BSs) constitute an important class of states
originally introduced for the quantum electromagnetic (e.m.) field
\cite{sto}. They are defined as a finite linear superposition of
field number states $\ket{n}$ weighted by a binomial counting
probability distribution. BSs are characterized by a maximum number
of excitations $N$ and a probability of single excitation occurrence
$p$. When they are also characterized by a mean phase $\phi$
\cite{vid1}, they are called generalized binomial states (GBSs)
\cite{sto,lof1}. One of their peculiarities is that they are
``intermediate'' between the number state and the coherent state.
Because of their interesting features, BSs have been subject
of several studies aimed at determining their properties and
possible applications
\cite{sto,vid1,vid2,dat,joshi1,joshi2,lee,abdel-aty}. Recently,
interest has again arisen about GBSs because of the discover that
they can be exploited as reference states within schemes devoted at measuring the
canonical phase of quantum electromagnetic fields
\cite{pregnell1,pregnell2}. Moreover, in the context of cavity
quantum electrodynamics (CQED), GBSs of e.m. field appear to be
generated quite naturally and then used for fundamental or
applicative purposes. In fact, they have been shown to be
efficiently produced, within the current experimental capabilities
\cite{raimond,harochebook}, by standard atom-cavity interactions
exploiting two-level atoms crossing one at a time a high-$Q$ cavity
\cite{lof2,lof3}. A coherent orthogonal superposition of two GBSs of e.m. field is also a good candidate to
study the classical-quantum border, due to the fact that each component of the superposition presents a non-zero
mean field \cite{lof3}.

However, most of the properties of BSs and GBSs have been previously obtained
by algebraic procedures \cite{sto,vid1,lof1,vid2,dat,fu,fan}, that
often make non intuitive both the determination and interpretation of these
properties. In particular, the
orthogonality between two GBSs, that plays a fundamental rule in the study of classical-quantum superpositions, has been reported only recently \cite{lof1}.
GBSs have been also shown to be eigenstates of an eigenvalue
equation using the Holstein-Primakoff realization of the Lie algebra
$SU(2)$ \cite{fu}. This has led to the observation that they can be
viewed as special $SU(2)$ coherent states \cite{fu,fan}. It thus
appears of interest to find simple and physically intuitive methods
enabling us to get known and new properties of GBSs. This paper
addresses this issue with reference to the GBSs of e.m. field.

To this end, we begin by noting that, due to the presence of a maximum number of excitations (photons) $N$, GBSs of e.m. field
recall collective states describing excitations in systems of $N$ atoms. We wish to
show that this analogy can be put on solid grounds
establishing a bijective mapping between the binomial states of $N$
photons and the well-known coherent atomic states (CASs) of $N$
two-level (spin-like) atoms \cite{arecchi,mandelbook}. Therefore, our target is to prove that
GBSs are the electromagnetic correspondent of CASs. This
correspondence aims at bringing at light properties of the GBSs by exploiting analogous properties possessed by the CASs. In particular, on the basis of a convenient geometrical representation, we succeed in interpreting GBSs as pseudo-angular momentum states. In order to obtain more knowledge about non-classical characteristics of GBSs, we finally analyze their
squeezing behavior.

The paper is organized as follows. In Section~\ref{GBS} we give the
definition of $N$-photon generalized binomial state ($N$GBS) and
some of its principal properties. In Section~\ref{analogy} we show
the bijection between $N$GBS and CAS, providing then an appropriate
angular momentum operator approach for them. In
Section~\ref{GBSbasis}, using this correspondence, we construct both
an over-complete basis of $N$GBSs and an orthonormal basis of e.m.
field states analogous to the rotated Dicke states. In
Section~\ref{squeezing} we analyze the squeezing behavior of the
$N$GBSs and in Section~\ref{conclusion} we report our conclusive
discussions.

\section{\label{GBS}Generalized binomial states of e.m. field}
The $N$-photon generalized binomial state ($N$GBS) is defined as
\cite{sto,vid1,lof2,lof3}
\begin{equation}
\ket{N,p,\phi}=\sum_{n=0}^N\left[{N\choose
n}p^{n}(1-p)^{N-n}\right]^{1/2}e^{in\phi}\ket{n},\label{NGBS}
\end{equation}
it is normalized and characterized by the probability of single
photon occurrence $0\leq p\leq1$ and the mean phase $\phi$. Note
that the $N$GBS reduces to the vacuum state $\ket{0}$ when $p=0$ and
to the number state $\ket{N}$ when $p=1$. In the particular case
where $\phi=0$, the $N$GBS of Eq.~(\ref{NGBS}) is simply named
``binomial state'' \cite{sto}. On the other hand, for
$N\rightarrow\infty$ and $p\rightarrow0$ in such a way that $Np=\textrm{cost}\equiv|\alpha|^2$, the
$N$GBS becomes the Glauber coherent state
$\ket{|\alpha|e^{i\phi}}$. Thus, as well-known a $N$GBS
interpolates between the number and the coherent state and presents
in general non-zero expectation values of the fields.

A relevant property for discussing the quantum-classical border is that, for a given $N$GBS it is possible to find a $N$GBS orthogonal to the first one. This is intuitively not obvious and, for example, it is a property not shared by coherent states. Orthogonality of two $N$GBSs has been recently reported using algebraic methods \cite{lof1} and it has been
shown that the inner product of two different $N$GBSs,
$\ket{N,p,\phi}$ and $\ket{N,p',\phi'}$, is given by
\begin{eqnarray}
\scal{N,p,\phi}{N,p',\phi'}=\sum_{n=0}^N{N\choose
n}(pp')^{n/2}[(1-p)(1-p')]^{(N-n)/2}e^{in(\phi'-\phi)},\label{scabin}
\end{eqnarray}
it thus vanishes when $p'=1-p$ and $\phi'=\pi+\phi$, that is
\cite{lof1}
\begin{equation}\label{GBSorthogonality}
\scal{N,p,\phi}{N,1-p,\pi+\phi}=0.
\end{equation}

In the following, we shall exploit another important aspect of the
$N$GBSs, that is the fact that they are the electromagnetic correspondent of the
well-known coherent atomic states (CASs) \cite{arecchi}. Such a correspondence shall permit us to
easily translate the properties of the CASs to the $N$GBSs.

\section{\label{analogy}Correspondence between $N$-photon generalized binomial states and coherent atomic states}
\begin{figure}
\begin{center}
\resizebox{0.3\columnwidth}{0.18\textheight}{%
\includegraphics{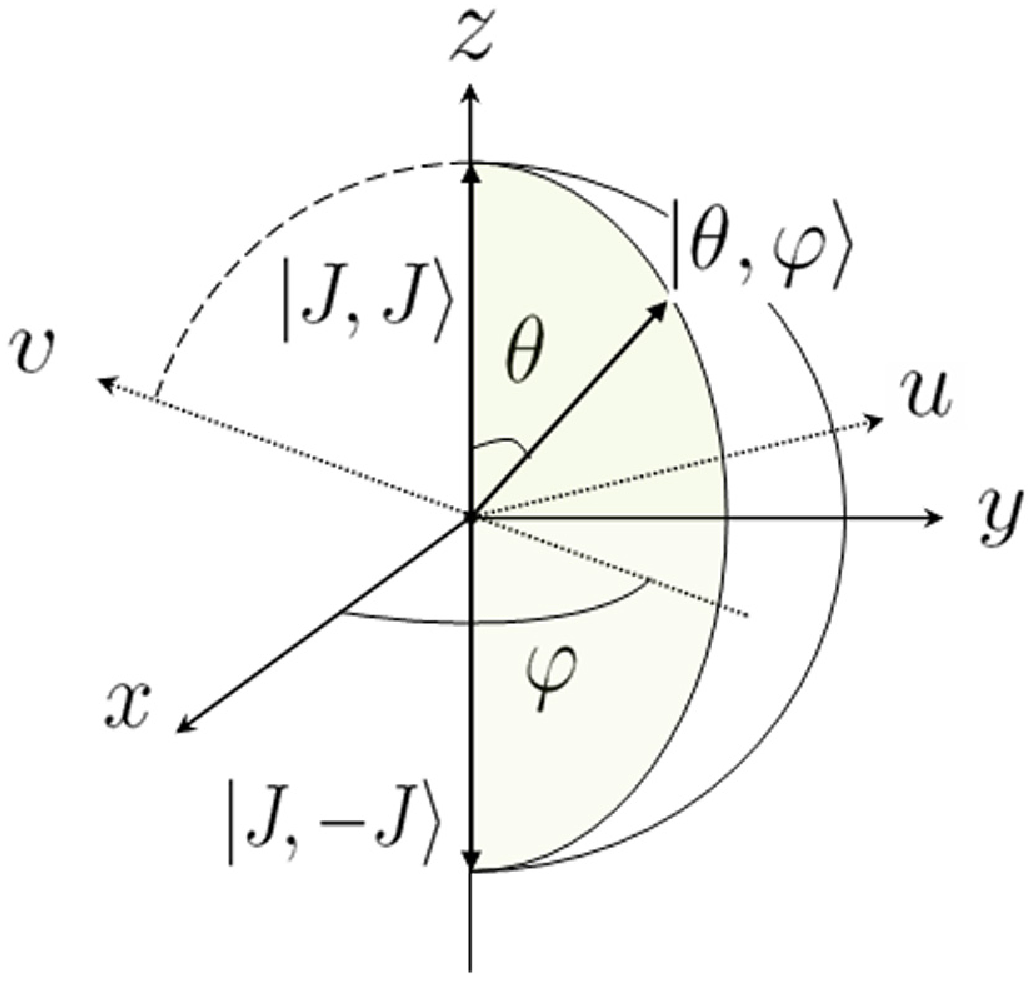}}
\hspace{1.8cm}
\resizebox{0.3\columnwidth}{0.18\textheight}{%
\includegraphics{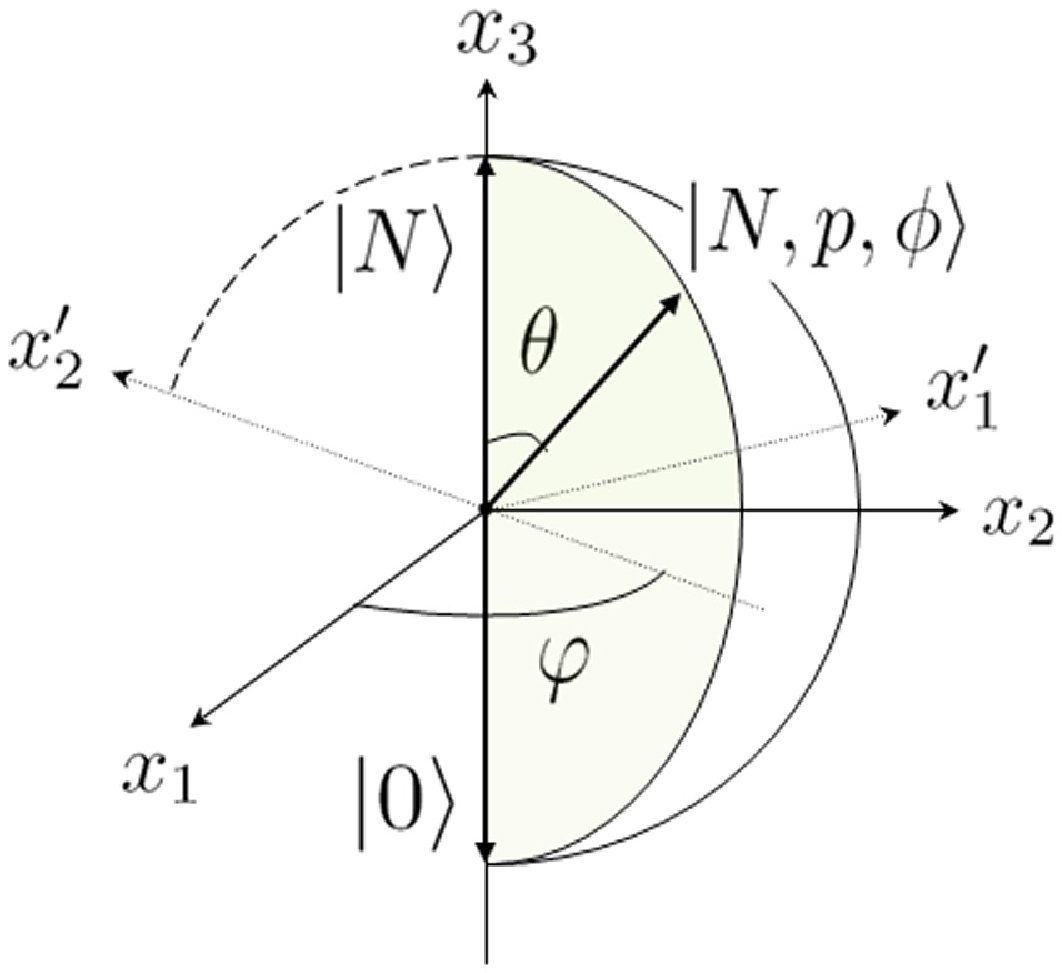}}
\caption{Geometrical representations of the coherent atomic state
(CAS) $\ket{\theta,\varphi}$ (left figure) and of the $N$-photon
generalized binomial state ($N$GBS) $\ket{N,p,\phi}$ (right figure).
The total angular momentum of the CAS points along the direction
$(\theta,\varphi)$. The probability of single photon occurrence $p$
and the mean phase $\phi$ of the $N$GBS are linked to the polar
angle $\theta$ and the azimuthal angle $\varphi$ according to the
relations $p\equiv\cos^2(\theta/2)$ and $\phi\equiv2\pi-\varphi$.}
\label{fig:CASGBS}
\end{center}
\end{figure}

A coherent atomic state (CAS) is a particular collective state of
$N$ two-level atoms that can be cast in the form \cite{arecchi}
\begin{equation}\label{CASdef}
\ket{\theta,\varphi}=\sum_{n=0}^{2J}\left[{2J\choose
n}\left(\cos^2\frac{\theta}{2}\right)^n\left(\sin^2\frac{\theta}{2}\right)^{2J-n}\right]^{1/2}e^{-in\varphi}\ket{J,-J+n},
\end{equation}
where $\ket{J,-J+n}$ is the Dicke state of collective angular
momentum $J$ and excitation $n$ with $0\leq J\leq N/2$ if $N$ is
even or $1/2\leq J\leq N/2$ if $N$ is odd \cite{mandelbook,dicke}.
It is well-known that the CAS of Eq.~(\ref{CASdef}) can be
obtained by rotation of the highest excitation Dicke state
$\ket{J,J}$ through an angle $\theta$ about an axis
$\hat{\mathbf{u}}=(-\sin\varphi,\cos\varphi,0)$ on the plane
$(x,y)$, as shown in Appendix~\ref{CAS} and illustrated in the Bloch-sphere of Fig.~\ref{fig:CASGBS}. Thus, indicating by $R_{\theta,\varphi}$ the
rotation operator $\exp\{-i\theta\mathbf{J}\cdot\hat{\mathbf{u}}\}$, we have $\ket{\theta,\varphi}\equiv
R_{\theta,\varphi}\ket{J,J}$.

Glauber coherent states are usually considered the electromagnetic analogous of the coherent atomic states, because of
common characteristics such as the formal generation by means of a displacement operator, the over-completeness property and the effective production by a classical source \cite{mandelbook}. However, relevant differences occur between field coherent states and CASs. For example, for a fixed $J$, it is possible to find two orthogonal CASs $\ket{\theta,\varphi}$ and $\ket{\pi-\theta,\pi+\varphi}$ (see Appendix~\ref{CAS}), while two different coherent states are never orthogonal. Moreover, a CAS is represented by a finite superposition of Dicke states while a coherent state is represented by an infinite superposition of Fock states. In this section we show that the CASs have indeed an exact electromagnetic equivalent in the $N$GBSs.

By comparing the form of the $N$GBS given in Eq.~(\ref{NGBS}) with that one of the CAS given in
Eq.~(\ref{CASdef}), it is readily seen that the two states assume exactly the same form if one associates the Dicke states $\ket{J,-J+n}$ to the number states $\ket{n}$ and sets the parametric correspondences
\begin{equation}\label{parametricCASGBS}
N\equiv 2J,\ p\equiv\cos^2(\theta/2),\
\phi\equiv2\pi-\varphi\quad\Rightarrow\quad\ket{N,p,\phi}\leftrightarrow\ket{\theta,\varphi}.
\end{equation}
This shows that the $N$-photon generalized binomial states can be put in a bijective mapping with the coherent atomic
states. This permits us to easily transfer the properties of the CASs to the $N$GBSs. In particular, the $N$GBS may be
geometrically represented in a pseudo Bloch-sphere as in Fig.~\ref{fig:CASGBS} and formally obtained by a rotation of the highest number state $\ket{N}$.

This geometrical representation of the $N$GBS, together with the
parametric associations of Eq.~(\ref{parametricCASGBS}), allows to
immediately obtain its limit properties, that is for $\theta=0$
($p=1$) the $N$GBS becomes the number state $\ket{N}$ while for
$\theta=\pi$ ($p=0$) it becomes the vacuum (ground) state $\ket{0}$.
Moreover, the orthogonality property of two $N$GBSs can be immediately
determined from the corresponding one of the CASs, where two states are orthogonal if they point in opposite directions.
$\scal{\theta,\varphi}{\pi-\theta,\pi+\varphi}=0$, and deduced from
Fig.~\ref{fig:CASGBS}. In fact, the state $\ket{N,p',\phi'}$
``antiparallel'' to $\ket{N,p,\phi}$ is described by the angles
$\theta'=\pi-\theta$ and $\varphi'=\pi+\varphi$, which correspond
respectively to the $N$GBS parameters $p'=1-p$ and $\phi'=\pi+\phi$,
as seen from Eq.~(\ref{parametricCASGBS}). This result gives
$\scal{N,p,\phi}{N,1-p,\pi+\phi}=0$ as previously given by
Eq.~(\ref{GBSorthogonality}). In particular, one obtains that there is one and only one orthogonal state to a $N$GBS which is again a $N$GBS.

In order to follow our analogy, we now have to choose appropriate
non-rotated angular momentum operators acting on the number (Fock)
states analogous to the non-rotated atomic angular momentum
operators $J_z,J_\pm$ and then construct the rotated operators. The
choice naturally goes to the well-known Holstein-Primakoff
realization of Lie algebra $SU(2)$ \cite{holstein,fu,palev}, which
utilizes boson creation and annihilation operators $a^\dag,a$ and
defines the pseudo-angular momentum operators acting on a $N$GBS in the following way:
\begin{equation}\label{holsteinoperators}
J_3^{(N)}=a^\dag a-N/2,\quad J_+^{(N)}=a^\dag\sqrt{N-a^\dag a},\quad
J_-^{(N)}=\sqrt{N-a^\dag a}a.
\end{equation}
The operators of Eq.~(\ref{holsteinoperators}) are the analogous of
the non-rotated angular momentum operators given in
Eq.~(\ref{collect spin op}) of Appendix~\ref{CAS} for the collective system of $N$
two-level atoms and satisfy the same commutation rules
\begin{equation}\label{commutation holstein}
[J_+^{(N)},J_-^{(N)}]=2J_3,\quad[J_3^{(N)},J_\pm^{(N)}]=\pm
J_\pm^{(N)},\quad[J_i^{(N)},(J^{(N)})^2]=0.
\end{equation}
The action of these operators on the number state $\ket{n}$ is then
\begin{eqnarray}
J_3^{(N)}\ket{n}&=&(n-N/2)\ket{n},\quad
J_+^{(N)}\ket{n}=\sqrt{(N-n)(n+1)}\ket{n+1},\nonumber\\
J_-^{(N)}\ket{n}&=&\sqrt{(N-n+1)n}\ket{n-1},
\end{eqnarray}
from which we readily obtain the eigenvalues equations
\begin{equation}
J_+^{(N)}\ket{N}=J_-^{(N)}\ket{0}=0,\quad
J_3^{(N)}\ket{N}=(N/2)\ket{N},\quad J_3^{(N)}\ket{0}=(-N/2)\ket{0}.
\end{equation}
In analogy with the case of CAS, the rotation operator
$R_{\theta,\varphi}^{(N)}=\exp\{-i\theta\mathbf{J}\cdot\hat{\mathbf{x}}'_1\}$
that produces a rotation through an angle $\theta$ about an axis
$\hat{\mathbf{x}}'_1=(-\sin\varphi,\cos\varphi,0)$ on the plane
$(x_1,x_2)$, as illustrated in Fig.~\ref{fig:CASGBS}, results to be
\begin{equation}
R_{\theta,\varphi}^{(N)}=e^{-\eta J_+^{(N)}+\eta^\ast J_-^{(N)}},
\end{equation}
where $\eta=(\theta/2)e^{-i\varphi}$. The $N$-photon generalized
binomial state $\ket{N,p,\phi}$ is then obtained by rotation of the
number state $\ket{N}$ taking into account the parametric associations given by
Eq.~(\ref{parametricCASGBS}). In other words, we have
\begin{equation}
\ket{N,p,\phi}=R_{\theta,\varphi}^{(N)}\ket{N}=e^{-\eta
J_+^{(N)}+\eta^\ast J_-^{(N)}}\ket{N}.
\end{equation}

It is now very simple to find the rotated pseudo-angular momentum
operators $J_3^{\prime(N)},J^{\prime(N)}_\pm$ for the $N$GBSs. In
fact, using the expressions of the rotated operators for the
collective atomic system given in Eqs.~(\ref{rotatedJz}) and
(\ref{rotatedJpm}) of Appendix~\ref{CAS} together with the
parametric correspondences of Eq.~(\ref{parametricCASGBS}), we have
\begin{eqnarray}\label{rotatedoperatorGBS}
J_3^{\prime(N)}&=&(2p-1)J_3^{(N)}+\sqrt{p(1-p)}[e^{i\phi}J_+^{(N)}+e^{-i\phi}J_-^{(N)}],\nonumber\\
J^{\prime(N)}_+&=&[J^{\prime(N)}_-]^\dag=e^{-i\phi}[pe^{i\phi}J_+^{(N)}-(1-p)e^{-i\phi}J_-^{(N)}-2\sqrt{p(1-p)}J_3^{(N)}],
\end{eqnarray}
from which we obtain the expected eigenvalues equations
\begin{eqnarray}
J_3^{\prime(N)}\ket{N,p,\phi}&=&(N/2)\ket{N,p,\phi},\quad
J_3^{\prime(N)}\ket{N,1-p,\pi+\phi}=-(N/2)\ket{N,p,\phi},\nonumber\\
J^{\prime(N)}_+\ket{N,p,\phi}&=&J^{\prime(N)}_-\ket{N,1-p,\pi+\phi}=0.
\end{eqnarray}
In this sense, the $N$GBS $\ket{N,p,\phi}$ defined in
Eq.~(\ref{NGBS}) can be viewed as a pseudo-angular momentum state
corresponding to the maximum value of the angular momentum $J=N/2$
along a direction $(\theta,\varphi)$ in a Bloch sphere, where the
polar and azimuthal angles are fixed by the binomial state
parameters as $\theta=2\arccos\sqrt{p}$ and $\varphi=2\pi-\phi$. A similar form of the operator $J_3^{\prime(N)}$ given in Eq.~(\ref{rotatedoperatorGBS}) was previously obtained by using an algebraic ladder operator approach \cite{fu}.

Analogously to the case of two CASs, two different $N$GBSs are linked by a rotation operator
$T_{pp'\phi\phi'}$ in the pseudo Bloch-sphere as
\begin{equation}
\ket{N,p',\phi'}=T_{pp'\phi\phi'}\ket{N,p,\phi}.
\end{equation}
The form of $T_{pp'\phi\phi'}$ is given by Eqs.~(\ref{CASCASoperator}) and (\ref{newangles})
of Appendix~\ref{CAS} for the CASs case and using
the parametric correspondences of Eq.~(\ref{parametricCASGBS}).

\section{\label{GBSbasis}$N$-photon generalized binomial states basis}
In this section we shall focus our attention on an aspect that has not been previously discussed, that is the possibility to
construct a basis of e.m. field states using $N$GBSs. The analogy
with the CASs indicates both how to immediately obtain an
over-complete basis of $N$GBSs and how to construct an orthonormal basis
of pseudo-angular momentum states with two orthogonal $N$GBSs
representing the ground and highest angular momentum state.

\subsection{$N$GBSs over-complete basis}
It is well-known that the CASs form an over-complete basis in the
Hilbert space spanned by the $2J+1$ Dicke states
$\ket{J,-J},\ket{J,-J+1},\ldots,\ket{J,J}$, as described in detail
in Appendix~\ref{CAS} \cite{arecchi,mandelbook}. As a consequence, the $N$GBSs can be also shown to form
an over-complete basis in the space spanned by the $N+1$
number states $\ket{0},\ket{1},\ldots,\ket{N}$. In fact, by
exploiting Eqs.~(\ref{CASunity}) and (\ref{parametricCASGBS}) of Appendix~\ref{CAS} we
get the completeness relation
\begin{equation}\label{GBSunity}
(N+1)\int\frac{\textrm{d}\Omega}{4\pi}\ket{N,p,\phi}\bra{N,p,\phi}=1,
\end{equation}
where the infinitesimal solid angle d$\Omega$ can be written in
terms of the $N$GBS parameters $p,\phi$ as
\begin{equation}\label{solidangleGBS}
\textrm{d}\Omega=\sin\theta\textrm{d}\theta\textrm{d}\varphi=-2\sqrt{1-p}\
\textrm{d}p\textrm{d}\phi.
\end{equation}
The expansion of an arbitrary e.m. field state
$\ket{\psi}=\sum_n^Nc_n\ket{n}$ in terms of $N$GBSs becomes
\begin{equation}\label{GBSbasis2}
\ket{\psi}=(N+1)\int\frac{\textrm{d}\Omega}{4\pi}\frac{A(\tau^\ast)}{[1+|\tau|^2]^{N/2}}\ket{N,p,\phi},
\end{equation}
where $\tau=e^{i\phi}\sqrt{(1-p)/p}$ and the amplitude function
$A(\tau^\ast)$ is given by
\begin{equation}\label{factorA}
A(\tau^\ast)\equiv\sum_{n=0}^Nc_n{N\choose
n}^{1/2}(\tau^\ast)^n=[1+|\tau|^2]^{N/2}\scal{N,p,\phi}{\psi}.
\end{equation}
These results indicate that the $N$GBSs may be used to construct a new
representation for non-classical e.m. field states.

\subsection{Complete orthonormal basis}
The equivalence between CASs and $N$GBSs has permitted us to easily
construct the rotated angular momentum operators
$J_3^{\prime(N)},J^{\prime(N)}_\pm$ given in
Eq.~(\ref{rotatedoperatorGBS}). We have also seen that the two
orthogonal $N$GBSs $\ket{N,p,\phi}$, $\ket{N,1-p,\pi+\phi}$, related
to direction $(\theta,\varphi)$ in Fig.~\ref{fig:CASGBS}, are the
eigenstates of the operator $J_3^{\prime(N)}$ with eigenvalues $N/2$
and $-N/2$, respectively. Thus, they represent the highest and
ground state of the orthonormal basis formed by all the eigenstates
of the operator $J_3^{\prime(N)}$. It is then possible to obtain
the expression of these e.m. field states, each orthogonal to the
two orthogonal $N$GBSs, that represent the electromagnetic
analogous of the rotated Dicke states.

This goal can be reached by repeatedly applying the rotated raising
operator $J^{\prime(N)}_+$ of Eq.~(\ref{rotatedoperatorGBS}) on the
ground state $\ket{N,1-p,\pi+\phi}$. In fact, because of the
commutation rules given in Eq.~(\ref{commutation holstein}), we also
have that the operator $J^2$ commutes with $J^{\prime(N)}_+=0$, so
that the action of the rotated raising operator is inside the
subspace $J=N/2$. We then obtain the orthonormal basis of
e.m. field states
$\mathcal{B}=\{\ket{\Delta_{p,\phi}(N/2,m-N/2)},m=0,1,\ldots,N\}$, which we call ``Delta'' states,
defined as
\begin{eqnarray}\label{Gammastates}
\ket{\Delta_{p,\phi}(N/2,m-N/2)}&\equiv&{N\choose
m}^{-1/2}\frac{[J^{\prime(N)}_+]^m}{m!}\ket{N,1-p,\pi+\phi}\nonumber\\
&=&\frac{J^{\prime(N)}_+}{\sqrt{m(N-m+1)}}\ket{\Delta_{p,\phi}(N/2,m-1-N/2)}
\end{eqnarray}
where we have set
$\ket{\Delta_{p,\phi}(N/2,-N/2)}\equiv\ket{N,1-p,\pi+\phi}$ and
$\ket{\Delta_{p,\phi}(N/2,N/2)}\equiv\ket{N,p,\phi}$.

As an example, let us consider the case $N=2$. The three Delta
states forming the basis are the two orthogonal 2GBSs
$\ket{2,p,\phi},\ket{2,1-p,\pi+\phi}$, given by Eq.~(\ref{NGBS}),
and the intermediate state $\ket{\Delta_{p,\phi}(1,0)}$ obtained by
Eq.~(\ref{Gammastates}) and having the form
\begin{equation}\label{2Gamma}
\ket{\Delta_{p,\phi}(1,0)}=\sqrt{2p(1-p)}\ket{0}+(2p-1)e^{i\phi}\ket{1}-\sqrt{2p(1-p)}e^{i2\phi}\ket{2}.
\end{equation}
We point out that the previous Delta state arises in the process of
generation of entangled 2GBSs in two separate cavities and in
principle it can be conditionally obtained during that process
\cite{lof4}.

Therefore, the analogy between coherent atomic states and $N$-photon
generalized binomial states has allowed us to introduce a new class
of e.m. field states that we have named ``Delta'' states which are
the analogous of rotated Dicke states.

\section{\label{squeezing}Squeezing of $N$-photon generalized binomial states}
In this section we shall analyze the squeezing behavior of the
$N$-photon generalized binomial states, in order to characterized some of their nonclassical properties.

The squeezing of a single-mode e.m. field state are usually examined
taking into account the two field quadratures $a_X,a_P$ defined as
\cite{mandelbook}
\begin{equation}\label{quadratures}
a_X=a+a^\dag,\quad a_P=\frac{a-a^\dag}{i},
\end{equation}
where $a,a^\dag$ are respectively the annihilation and creation
operators of the field mode. The two quadratures $a_X,a_P$ behave like dimensionless
canonical conjugates, satisfying the following commutation and
uncertainty relations for any field state:
\begin{equation}\label{quadraturecommutation}
[a_X,a_P]=2i,\quad \ave{\Delta a_X^2}\ave{\Delta a_P^2}\geq1,
\end{equation}
where the dispersion $\ave{\Delta a_K^2}$ ($K=X,P$) is given by
$\ave{\Delta a_K^2}=\ave{a_K^2}-\ave{a_K}^2$. A squeezed
state is defined as a state for which either $a_X$ or $a_P$ has a
dispersion below unity with a corresponding increase in the
dispersion of the other quadrature \cite{mandelbook}. In order to investigate the
squeezing behavior of $N$GBSs, that are not states of minimum uncertainty for the quadratures, it is convenient to introduce the squeezing indexes $S_X,S_P$ as
\begin{equation}\label{squeezingindex}
S_K=1-\ave{\Delta a_K^2},\qquad(K=X,P)
\end{equation}
so that squeezing occurs when $S_K>0$.

\begin{figure}
\begin{center}
\resizebox{0.33\columnwidth}{0.18\textheight}{%
\includegraphics{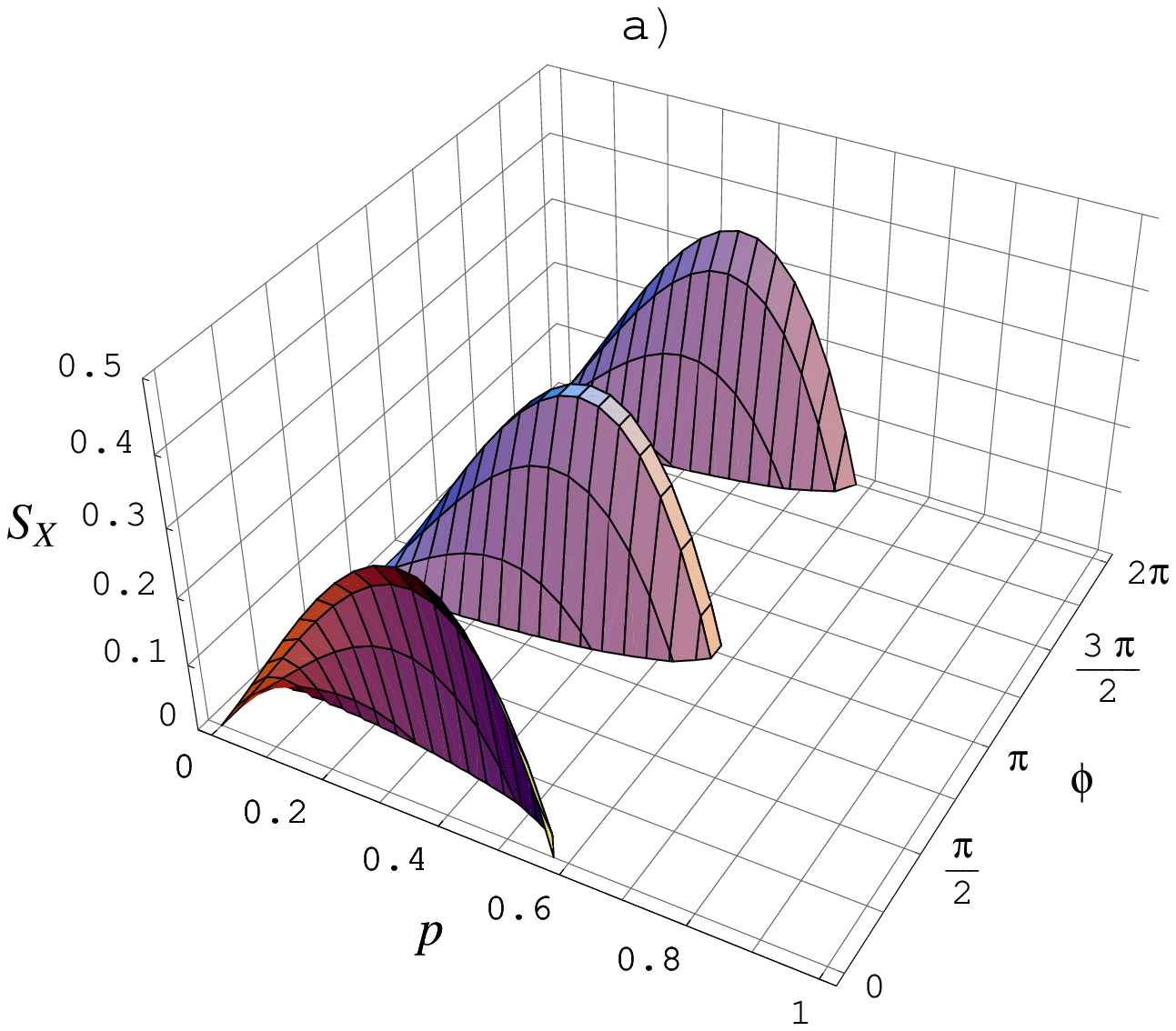}}
\hspace{1cm}
\resizebox{0.33\columnwidth}{0.18\textheight}{%
\includegraphics{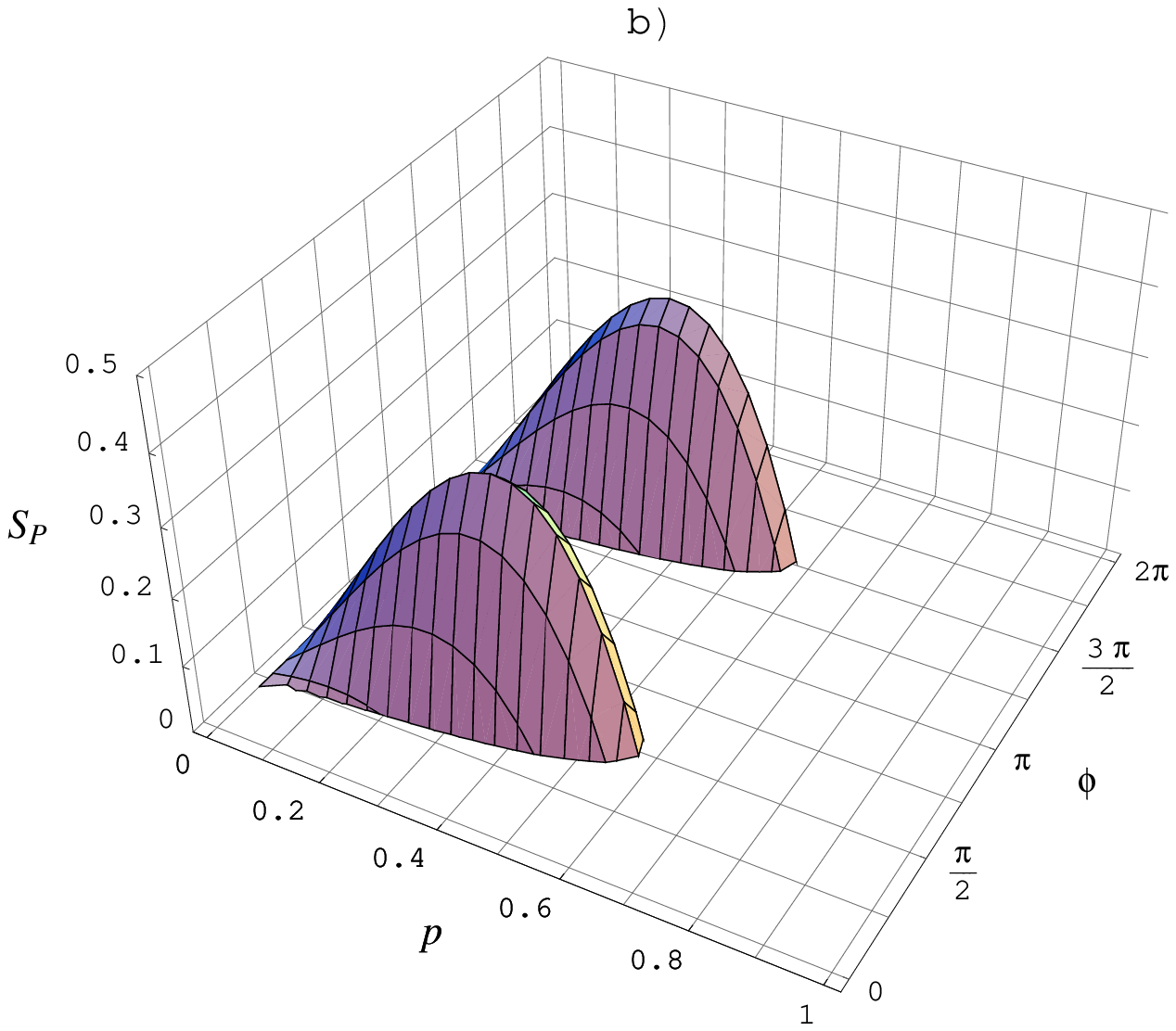}}\\${}$\\
\resizebox{0.33\columnwidth}{0.18\textheight}{%
\includegraphics{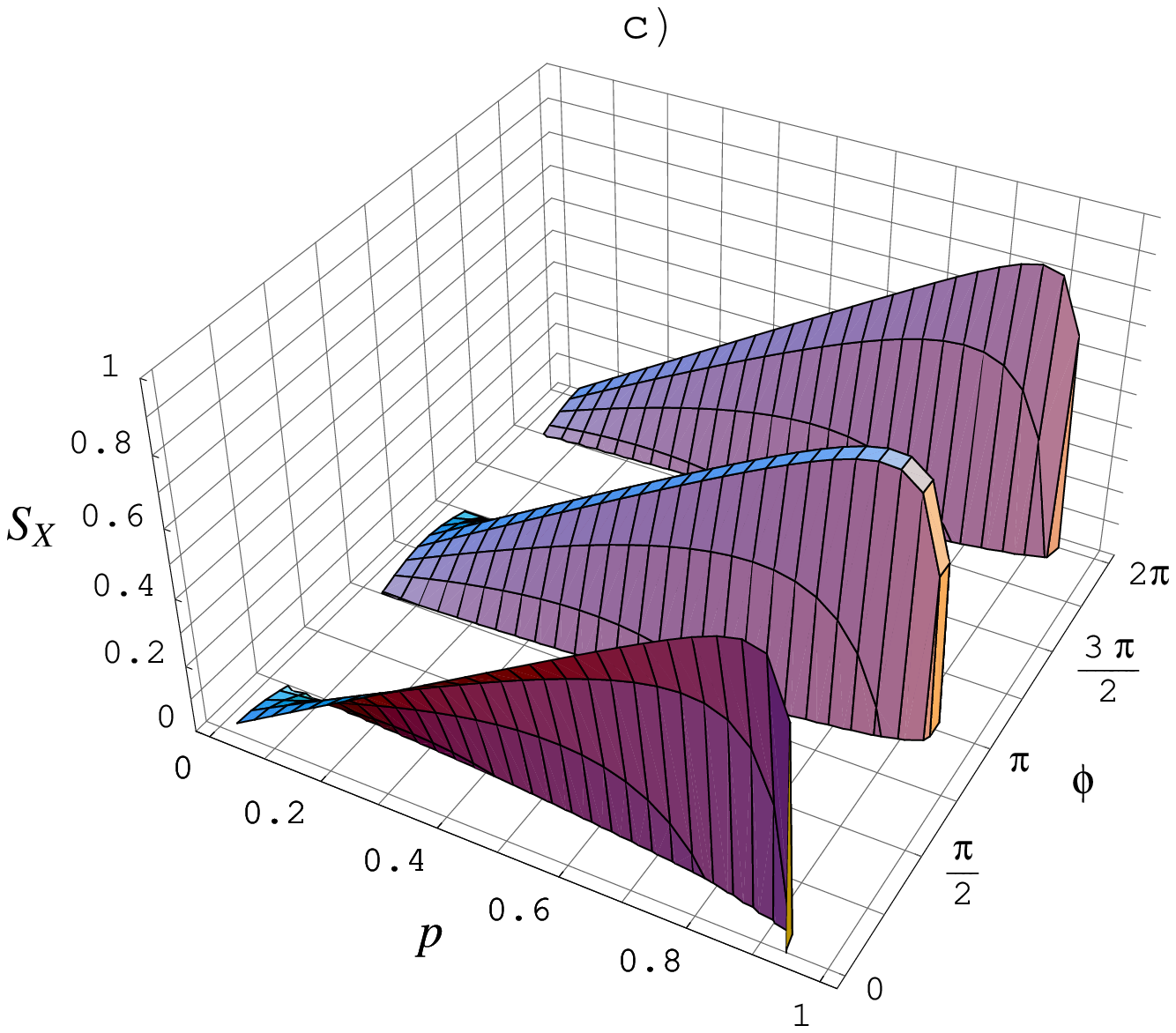}}
\hspace{1cm}
\resizebox{0.33\columnwidth}{0.18\textheight}{%
\includegraphics{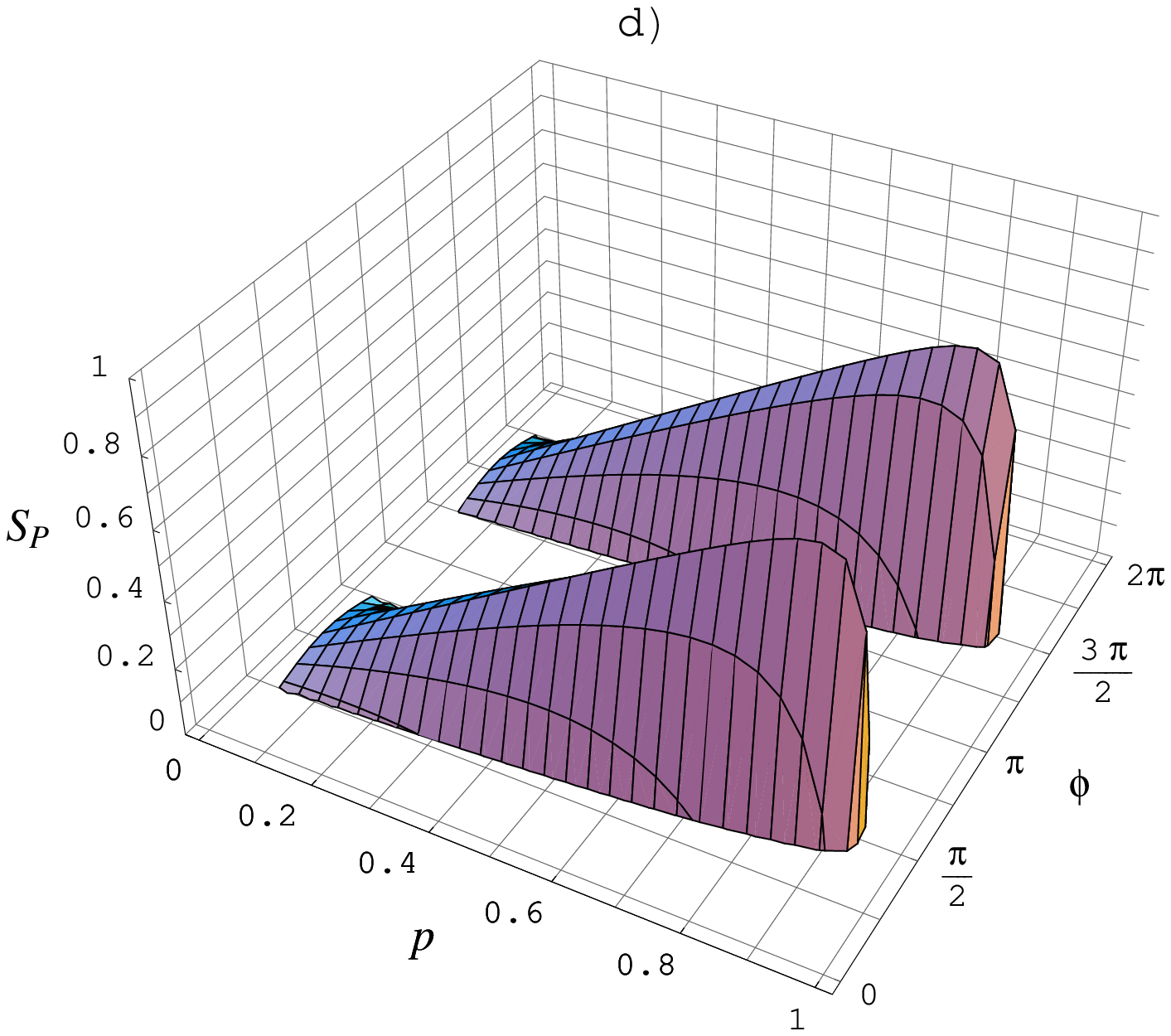}}
\caption{Squeezing indexes $S_X$ [plots (a) and (c)] and $S_P$
[plots (b) and (d)] as a function of the probability of single
photon occurrence $p$ and the mean phase $\phi$, for $N=2$ [plots
(a) and (b)] and $N=100$ [plots (c) and (d)]. Squeezing of the
quadratures $a_X$ or $a_P$ respectively occurs when $S_X>0$ or
$S_P>0$.} \label{fig:squeeze}
\end{center}
\end{figure}
Applying the previous considerations to the $N$GBS $\ket{N,p,\phi}$
defined in Eq.~(\ref{NGBS}) we obtain
\begin{eqnarray}\label{NGBSsqueeze}
S_X&=&-2Np-A(N,p)\cos2\phi+B^2(N,p)\cos^2\phi,\nonumber\\
S_P&=&-2Np+A(N,p)\cos2\phi+B^2(N,p)\sin^2\phi,
\end{eqnarray}
where we have set
\begin{eqnarray}
A(N,p)&=&2\sqrt{N(N-1)}p(1-p)\sum_{n=0}^{N-2}\left[{N\choose
n}{N-2\choose n}\right]^{1/2}p^n(1-p)^{N-2-n},\nonumber\\
B(N,p)&=&2\sqrt{Np(1-p)}\sum_{n=0}^{N-1}\left[{N\choose
n}{N-1\choose n}\right]^{1/2}p^n(1-p)^{N-1-n}.
\end{eqnarray}
The squeezing behavior is analyzed by fixing the value of the
maximum number of photons $N$ and varying the value of the
parameters $p,\phi$. We find that squeezing occurs, that is $S_X>0$
or $S_P>0$, for some values of $p,\phi$ at any value of $N\geq1$. We
also find that the larger is $N$, the more squeezed is the
quadrature $a_X$ or $a_P$. We represent this behavior in
Fig.~\ref{fig:squeeze}, where we plot the squeezing indexes $S_X$
and $S_P$ as functions of the probability of single photon
occurrence $p$ and the mean phase $\phi$, for the two values $N=2$
and $N=100$. We observe that, while the range of values of $\phi$
where squeezing occurs remain unchanged for different $N$, the
squeezing range of $p$ increases with $N$. We finally note that, when the quadrature $a_X$ is squeezed, the quadrature
$a_P$ is not squeezed and viceversa.

Therefore, we may conclude that the $N$-photon generalized binomial
states are particular squeezed states of the e.m. field.

\section{\label{conclusion}Conclusion}
In this paper we have shown that the $N$-photon generalized binomial
states ($N$GBSs) of e.m. field are the electromagnetic
correspondent of the coherent atomic states (CASs) describing a set
of $N$ two-level atoms. We have shown that this equivalence simply arises
from an association of the parameters characterizing the two kind of
states, that is $N$GBSs and CASs have the same structure
(Sec.~\ref{analogy}). This correspondence has then
permitted us to immediately apply the known results for the CASs to
the $N$GBSs.

Using this analogy, we have described the $N$GBSs as pseudo-angular
momentum states, representable in a Bloch sphere and obtainable by a
rotation in the space of the number state $\ket{N}$.
We have also found the appropriate rotated angular momentum operators for $N$GBSs by using
the Holstein-Primakoff operators as non-rotated angular momentum
operators acting on the number states. This has permitted to immediately
obtain an eigenvalue equation for the $N$GBSs
(Sec.~\ref{analogy}). Note that an eigenvalue equation for binomial
states was previously obtained by means of a ladder operator
approach \cite{fu}.

Exploiting the analogy with the CAS, we have then constructed an
over-complete basis of $N$GBSs in the Hilbert space spanned by the
first $N+1$ number states (Sec.~\ref{GBSbasis}), showing that an
arbitrary single-mode field state with a maximum photon number $N$
can be expressed in terms of this basis.

The bijective mapping between $N$GBS and CAS has allowed to construct an orthonormal basis of e.m. field
states that are eigenstates of the rotated pseudo-angular momentum
operator. In this basis, two orthogonal $N$GBSs constitute the ground and
highest angular momentum state. These states, named ``Delta'', are a new class of e.m. field states that correspond to
rotated Dicke states of a $N$ two-level atoms system (Sec.~\ref{GBSbasis}). Moreover, they may be generated by standard atom-cavity interactions in the CQED framework \cite{lof2}.

Finally, we have analyzed the squeezing properties of the $N$GBSs (Sec.~\ref{squeezing}).
We have found that the $N$GBSs are particular squeezed states of the e.m. field.
In fact, squeezing of the field quadratures occurs at some values of the $N$GBS
parameters for a given value of the maximum number of photons $N\geq1$.

In conclusion, the perfect correspondence with CASs has allowed us
to obtain immediately the principal properties and also new
characteristics of $N$GBSs. The $N$GBSs are thus the ``true''
electromagnetic analogous of the CASs. The usual correspondence between Glauber coherent
states and coherent atomic states may be then viewed as a consequence of the fact that Glauber coherent states are somehow a limit of the $N$GBSs (see Sec.~\ref{GBS}). The analogy with the CASs appears to be the underlying
reason why $N$GBSs are naturally generated by interactions between
two-level atoms and a quantum e.m. field initially in its vacuum
(ground) state, as shown by the feasible efficient generation
schemes recently proposed in the context of CQED \cite{lof2,lof3},
based on two-level atoms interacting one at a time with a high-$Q$
cavity. These generation methods have been used to reach a 2GBS. However, the analogy with CASs suggest that the generation of
$N$GBSs with $N>2$ is also possible in the CQED framework and this
will be treated elsewhere. Finally, in view of the correspondence between $N$GBS and CAS, it would be also of interest to
investigate if a $N$GBS can be produced in a single-shot run by the
interaction of $N$ two-level atoms in a collective coherent atomic
state with a quantum e.m. field.
\\${}$\\

\noindent \textbf{Acknowledgements}
\\
Two of the authors, G.C. and R.L.F., wish to thank Prof. T. Arecchi
for his suggestions to investigate the relation between $N$GBSs and CASs.

\appendix
\section{\label{CAS}Dicke states and Coherent atomic states}
In this Appendix we review the definition of coherent atomic state
(CAS) and give some of its properties.

Let us consider $N$ identical two-level (spin-like) atoms described
by usual spin operators $s_i^{j}=\sigma_i^{j}/2$ ($i=x,y,z$;
$j=1,2,\ldots,N$), where $\sigma_i^{j}$ is the $i$-th Pauli matrix
relating to the $j$-th atom, or by the atomic lowering and raising
operators $\sigma_-^j=\ket{g_j}\bra{e_j}$,
$\sigma_+^j=\ket{e_j}\bra{g_j}$, with $g$ and $e$ representing
respectively the ground and excited state of the two-level atom. The
collective system of these $N$ atoms has the corresponding Hilbert
space spanned by the set of $2^N$ product states
\begin{equation}\label{product state}
\ket{\Phi_{k_1k_2\cdots
k_N}}=\prod_{j=1}^N\ket{k_j}.\quad(k_j=g_j,e_j)
\end{equation}
It is then convenient to introduce collective spin operators as
\begin{equation}\label{collect spin op}
J_i\equiv\sum_{j=1}^Ns_i^{j},\quad
J_\pm\equiv\sum_{j=1}^N\sigma_\pm^{j}=J_x\pm iJ_y,\quad
J^2=J_x^2+J_y^2+J_z^2,
\end{equation}
satisfying the commutation relations
\begin{equation}\label{commutation rules}
[J_+,J_-]=2J_z,\quad[J_z,J_\pm]=\pm J_\pm,\quad[J_i,J^2]=0.
\end{equation}
Note that the $J_+,J_-$ operators raise and lower the excitation of
the collective atomic system by unity in a manner such that this
excitation is distributed over all atoms. At this point, one
constructs the Dicke states $\ket{J,M}$ \cite{dicke}, which are
simply the usual angular momentum states, defined as the eigenstates
of $J_z,J^2$ with eigenvalues respectively $M,J(J+1)$ and given by
\begin{equation}\label{dickestate}
\ket{J,M=-J+n}\equiv\frac{1}{n!}{2J\choose
n}^{-1/2}J_+^n\ket{J,-J},\quad(n=0,1,\ldots,2J)
\end{equation}
where the ground state $\ket{J,-J}$ and the highest excitation state
$\ket{J,J}$ are defined by $J_-\ket{J,-J}=J_+\ket{J,J}=0$, with
$\textrm{min}|M|\leq J\leq N/2$. When $J=N/2$ the number $n$ of
Eq.~(\ref{dickestate}) can be viewed as the atomic excitation, so
that the Dicke states represent states of definite atomic excitation
in which the excitation is however distributed among the different
atoms. In this sense, the Dicke states of the $N$-atom system can be
considered as analogous to the Fock states of the e.m. field, the
atomic excitation being correspondent to the photon excitation
\cite{mandelbook}.

Let us now consider the rotation operator
$R_{\theta,\varphi}=\exp\{-i\theta\mathbf{J}\cdot\hat{\mathbf{u}}\}$
producing a rotation through an angle $\theta$ about an axis
$\hat{\mathbf{u}}=(-\sin\varphi,\cos\varphi,0)$ on the plane
$(x,y)$, as illustrated in Fig.~\ref{fig:CASGBS}. It is possible to
show that this operator can be factorized using the disentangling
theorem for angular momentum operators as \cite{arecchi}
\begin{equation}
R_{\theta,\varphi}=e^{-(\xi J_+-\xi^\ast J_-)}=e^{\tau^\ast
J_-}e^{-\ln(1+|\tau|^2)J_z}e^{-\tau J_+},
\end{equation}
where $\xi=(\theta/2)e^{-i\varphi}$ and
$\tau=\tan(\theta/2)e^{-i\varphi}$. In general, the action of this
rotation operator on an arbitrary atomic state $\ket{c}$ or an
atomic operator $Q$ reads like $\ket{c'}=R_{\theta,\varphi}\ket{c}$
and $Q'=R_{\theta,\varphi}QR_{\theta,\varphi}^{-1}$. The coherent
atomic state $\ket{\theta,\varphi}$ given in Eq.~(\ref{CASdef}) is
then obtained by rotation of the highest excitation Dicke state
$\ket{J,J}$, that is $\ket{\theta,\varphi}\equiv
R_{\theta,\varphi}\ket{J,J}$. This CAS can be equivalently generated
by rotation of the ground state $\ket{J,-J}$ through an angle
$\theta'=\pi-\theta$ about an axis $\hat{\mathbf{u}}'$ antiparallel
to the previous one $\hat{\mathbf{u}}$.

Note that for $\theta=0$ the CAS reduces to the Dicke state
$\ket{J,J}$ while for $\theta=\pi$ it reduces to the lowest
excitation (ground) Dicke state $\ket{J,-J}$. The coherent atomic
state $\ket{\theta,\varphi}$ is thus eigenstate of the rotated
angular momentum operator $J'_z$ with eigenvalue $J$, that is
$J'_z\ket{\theta,\varphi}=J\ket{\theta,\varphi}$, where
\begin{equation}\label{rotatedJz}
J'_z=J_z\cos\theta+\sin\theta(J_+e^{-i\varphi}+J_-e^{i\varphi})/2.
\end{equation}
Of course, the eigenvalue equation
$J^2\ket{\theta,\varphi}=J(J+1)\ket{\theta,\varphi}$ holds, as well.
At the same way, we can find the rotated raising and lowering
operators $J'_\pm$ which have the form
\begin{equation}\label{rotatedJpm}
J'_+=(J'_-)^\dag=e^{i\varphi}\left[J_+e^{-i\varphi}\cos^2(\theta/2)-J_-e^{i\varphi}\sin^2(\theta/2)-J_z\sin\theta\right].
\end{equation}
From Eqs.~(\ref{rotatedJpm}) and (\ref{CASdef}) one then obtains the
eigenvalue equation $J'_+\ket{\theta,\varphi}=0$, as expected.

It is also useful to have the global operator that links two
different CASs, that is
$\ket{\theta',\varphi'}=T_{\theta\theta'\varphi\varphi'}\ket{\theta,\varphi}$.
This operator can be found by observing that, from the definition of
CAS, we have
\begin{equation}
\ket{\theta',\varphi'}=R_{\theta',\varphi'}\ket{J,J}=R_{\theta',\varphi'}R_{\theta,\varphi}^{-1}\ket{\theta,\varphi},
\end{equation}
from which one obtains \cite{arecchi}
\begin{equation}\label{CASCASoperator}
T_{\theta\theta'\varphi\varphi'}\equiv
R_{\theta',\varphi'}R_{\theta,\varphi}^{-1}=\exp\left\{i\frac{\theta\theta'}{4}\sin(\varphi-\varphi')\right\}R_{\Theta,\Phi},
\end{equation}
where the angles $\Theta,\Phi$ are given by
\begin{equation}\label{newangles}
\Theta=[\theta^2+\theta'^2-2\theta\theta'\cos(\varphi-\varphi')]^{1/2},\quad
\tan\Phi=\frac{\theta'\sin\varphi'-\theta\sin\varphi}{\theta'\cos\varphi'-\theta\cos\varphi}.
\end{equation}

The coherent atomic states also form an over-complete set and they
can be exploited as basis. In fact, using the completeness of Dicke
states $\sum_n^{2J}\ket{J,-J+n}\bra{J,-J+n}=1$ it is found that
\cite{arecchi}
\begin{equation}\label{CASunity}
(2J+1)\int\frac{\textrm{d}\Omega}{4\pi}\ket{\theta,\varphi}\bra{\theta,\varphi}=1,
\end{equation}
and the expansion of an arbitrary atomic state
$\ket{c}=\sum_n^{2J}c_n\ket{J,-J+n}$ in terms of CASs reads like
\begin{equation}\label{CASbasis}
\ket{c}=(2J+1)\int\frac{\textrm{d}\Omega}{4\pi}\frac{f(\tau^\ast)}{[1+|\tau|^2]^J}\ket{\theta,\varphi},
\end{equation}
where the amplitude function $f(\tau^\ast)$ is given by
\begin{equation}\label{factorf}
f(\tau^\ast)\equiv\sum_n^{2J}c_n{2J\choose
n}^{1/2}(\tau^\ast)^n=[1+|\tau|^2]^J\scal{\theta,\varphi}{c}.\quad\left(\tau=\tan\frac{\theta}{2}e^{-i\varphi}\right)
\end{equation}

Two different CASs are not orthogonal in general, that is
$\scal{\theta,\varphi}{\theta',\varphi'}\neq0$. However, from the
geometrical representation of Fig.~\ref{fig:CASGBS} it is readily
seen that the CAS $\ket{\theta,\varphi}$ of Eq.~(\ref{CASdef})
admits an ``antiparallel'' orthogonal CAS given by
$\ket{\pi-\theta,\pi+\varphi}$, that is
$\scal{\theta,\varphi}{\pi-\theta,\pi+\varphi}=0$. In particular,
the scalar product of two CASs is given by \cite{arecchi,mandelbook}
\begin{equation}\label{CASproduct}
|\scal{\theta,\varphi}{\theta',\varphi'}|^2=[\cos(\Theta/2)]^{4J},
\end{equation}
where $\Theta$ is the angle between the two CASs.

\end{document}